\documentclass[usenatbib]{mn2e}
\usepackage{graphicx,times}

\newcommand\aj{AJ}
\newcommand\apj{ApJ}

\newcommand\apjs{ApJS}
\newcommand\mnras{MNRAS}
\newcommand\aap{A\&A}

\newcommand{\bea}{\begin{eqnarray}}
\newcommand{\eea}{\end{eqnarray}}

\newcommand{\Lx}{L_{\rm X}}
\newcommand{\Tx}{T_{\rm X}}
\newcommand{\Fx}{F_{\rm X}}

\newcommand{\rmag}{\>^{0.1}{\rm M}_r-5\log h}
\newcommand{\mpch}{\>h^{-1}{\rm {Mpc}}}
\newcommand{\msunh}{\>h^{-1}\rm M_\odot}

\newcommand{\etal}{{et al.~}}

\usepackage{color}

\title[Measuring the X-ray luminosities]{Measuring the X-ray luminosities
  of SDSS DR7 clusters from RASS} 

\author[Wang et al.]{\parbox[t]{\textwidth}{Lei Wang$^{1}$\thanks{E-mail:
      leiwang@pmo.ac.cn}, Xiaohu Yang$^{2,3}$\thanks{E-mail:
      xyang@sjtu.edu.cn}, Shiyin Shen$^{3}$, H. J. Mo$^{4}$, Frank C. van den
    Bosch$^{5}$, Wentao Luo$^{3}$, Yu Wang$^{6}$, Erwin T. Lau$^{7}$,
    Q.D. Wang$^{4}$, Xi Kang$^{1}$, Ran Li$^{8}$} \\
\vspace*{3pt} \\
$^1$ Purple Mountain Observatory, the Partner Group of MPI f\"{u}r Astronomie,
     2 West Beijing Road, Nanjing 210008, China\\
$^2$ Center for Astronomy and Astrophysics, Shanghai Jiao Tong University,
     Shanghai 200240, China\\
$^3$ Key Laboratory for Research in Galaxies and Cosmology, Shanghai
     Astronomical Observatory, Nandan Road 80, Shanghai 200030, China \\
$^4$ Department of Astronomy, University of Massachusetts, Amherst
     MA 01003-9305\\
$^5$ Department of Astronomy, Yale University, P.O. Box 208101,
     New Haven, CT 06520-8101, USA\\
$^6$ Key Laboratory for Research in Galaxies and Cosmology, Center
     for Astrophysics, University of Science and Technology of China, Hefei
     230026, China\\
$^7$ Department of Physics, Yale University, New Haven, CT 06520-8120, USA \\
$^8$ National Astronomical Observatories, Chinese Academy of Sciences, Beijing
     100871, China }

\begin{document}

\date{{\sc Draft: } \today }

\pagerange{\pageref{firstpage}--\pageref{lastpage}} \pubyear{2013}

\maketitle

\label{firstpage}


\begin{abstract}
  We use  ROSAT All Sky Survey  (RASS) broadband X-ray images  and the optical
  clusters identified from SDSS DR7  to estimate the X-ray luminosities around
  $\sim 65,000$ candidate clusters with masses $\ga 10^{13}\msunh$ based on an
  Optical to  X-ray (OTX) code  we develop. We  obtain a catalogue  with X-ray
  luminosity  for each  cluster. This  catalog contains  817 clusters  (473 at
  redshift $z\le 0.12$) with $S/N> 3$ in X-ray detection. We find about $65\%$
  of  these X-ray clusters  have their  most massive  member located  near the
  X-ray flux peak;  for the rest $35\%$, the most  massive galaxy is separated
  from the X-ray  peak, with the separation following  a distribution expected
  from a  NFW profile.   We investigate a  number of correlations  between the
  optical and  X-ray properties  of these X-ray  clusters, and find  that: the
  cluster X-ray luminosity is correlated with the stellar mass (luminosity) of
  the clusters, as  well as with the stellar mass  (luminosity) of the central
  galaxy and  the mass of the halo,  but the scatter in  these correlations is
  large.  Comparing  the properties of  X-ray clusters of similar  halo masses
  but having  different X-ray  luminosities, we find  that massive  halos with
  masses  $\ga  10^{14}\msunh$ contain  a  larger  fraction  of red  satellite
  galaxies when  they are brighter  in X-ray.  An  opposite trend is  found in
  central galaxies in relative  low-mass halos with masses $\la 10^{14}\msunh$
  where X-ray brighter clusters have smaller fraction of red central galaxies.
  Clusters  with masses  $\ga 10^{14}\msunh$  that are  strong  X-ray emitters
  contain  many more  low-mass satellite  galaxies than  weak  X-ray emitters.
  These results are also confirmed by checking X-ray clusters of similar X-ray
  luminosities but having different  characteristic stellar masses.  A cluster
  catalog containing the  optical properties of member galaxies  and the X-ray
  luminosity is available at {\it http://gax.shao.ac.cn/data/Group.html}.
\end{abstract}





\section{Introduction}


Clusters  of  galaxies  are  the   most  massive  virialized  objects  in  the
universe. Their  abundance and spatial distribution  are powerful cosmological
probes (e.g.,  Majumdar \& Mohr  2004; Vikhlinin et  al.  2009b; Mantz  et al.
2010a). In addition, galaxy clusters provide extreme environments for studying
the  formation  and  evolution  of   galaxies  within  the  framework  of  the
hierarchical build-up  of the most  massive halos.  One important  property of
clusters is that both their stellar and gas components are readily observable:
their gravitational wells are deep enough to retain energetic gas ejected from
their  member galaxies  which can  be observed  in optical  and  infrared. The
intracluster medium (ICM)  is also hot enough to be  observable in X-ray.  The
observed thermodynamic state of the  ICM is determined by the combined effects
of shock  heating during accretion,  radiative cooling, feedback  from stellar
evolution (stellar winds  and supernovae) and active galactic  nuclei, as well
as the magnetic fields, cosmic rays and turbulence.  The density, temperature,
and  entropy  profiles  of  the  ICM  therefore  carry  important  information
regarding the entire thermal history  of cluster formation.  The hot ICM, with
temperatures between $10^7$K and $10^8$K,  emits X-rays in the form of thermal
bremsstrahlung  and atomic line  emissions (e.g.,  Kellogg \etal  1971; Forman
\etal 1971).  By assuming hydrostatic equilibrium between the intracluster gas
and the cluster  potential, one can also derive the  gravitational mass of the
cluster using density and temperature measurements provided by X-ray data.

Clusters have also been observed by other means in addition to X-ray: optical,
infrared, radio,  Sunyaev-Zel'dovich effect and  gravitational lensing.  Among
these, the most complete cluster samples to date are optically-selected either
from  photometric  or spectroscopic  data.  Photometrically selected  complete
cluster samples can be constructed for  the most massive clusters and in large
redshift ranges. However, the properties  of their galaxy members are not well
understood.  In order  to have reliable membership assignments  of galaxies to
dark matter halos,  which is important for understanding  galaxy formation and
evolution in such systems, spectroscopic  data are needed. During the past two
decades,  numerous  group\footnote{In  this  paper,  we refer  to  systems  of
  galaxies as groups regardless of their richness, including isolated galaxies
  (i.e.,  systems  with  a  single  member) to  rich  clusters  of  galaxies.}
catalogues have  been constructed from  various redshift surveys  of galaxies,
most noticeably the CfA redshift survey (e.g.  Geller \& Huchra 1983), the Las
Campanas Redshift Survey (e.g.  Tucker  \etal 2000), the 2-degree Field Galaxy
Redshift  Survey (hereafter 2dFGRS;  Merch\'an \&  Zandivarez 2002;  Eke \etal
2004, Yang \etal 2005; Tago \etal 2006; Einasto \etal 2007), the high-redshift
DEEP2 survey  (Gerke \etal 2005), and  the Two Micron All  Sky Redshift Survey
(Crook \etal 2007).  Various group  catalogues have also been constructed from
redshift samples selected  from the Sloan Digital Sky  Survey (hereafter SDSS)
using different methods: friends-of-friends  (FOF) algorithm (e.g.  Goto 2005;
Merch\'an \&  Zandivarez 2005; Berlind  \etal 2006), the C4  algorithm (Miller
\etal 2005), and the halo-based  group finder (e.g., Yang \etal 2005; Weinmann
\etal 2006; Yang et al. 2007).  These catalogs provide galaxy groups that have
reliable galaxy memberships, which is important in probing the halo occupation
distribution (HOD)  statistics and galaxy  formation models (e.g. Yang  et al.
2008; 2009).

X-ray  selection  of galaxy  clusters  is reliable  but  typically  has a  low
efficiency.   Indeed,  even  if  survey  selections are  properly  taken  into
account, a significant  fraction of optically detected clusters  that obey the
scaling relation  between optical luminosity  and virial mass  (inferred from,
e.g.,  the velocity  dispersion of  member galaxies)  are undetected  in X-ray
(i.e., they  do not follow the  scaling relation between  X-ray luminosity and
virial mass).  This  has given rise to the notion that  there exists a genuine
population of  clusters that are  X-ray under-luminous (e.g.,  Castander \etal
1994; Lubin \etal 2004; Popesso \etal 2007; Castellano et al.  2011; Balogh et
al.  2011).  However, as will  be illustrated later, such an incompleteness in
X-ray cluster  detection is mainly due to  the blind search of  X-ray peaks in
relatively shallow observations. If  prior information about the positions and
sizes  of the  candidate  X-ray  clusters is  available,  the X-ray  detection
completeness can be improved dramatically.   In this paper, we study the X-ray
properties of  the optical clusters selected  by Yang et al.   (2007) from the
SDSS DR7 using the algorithm developed in Shen et al. (2008), which enables us
to obtain a much more complete X-ray cluster catalog than currently available.
The X-ray information so obtained adds to the wealth of optical information in
the SDSS  DR7 group catalog,  together providing a  useful data base  to study
galaxy formation and evolution in clusters of galaxies.

This paper  is organized  as follows.  In  Section \ref{sec_data},  we briefly
describe the group  samples to be used.  In  Section \ref{sec_OTX}, we outline
our Optical to X-ray (OTX) code used to estimate the X-ray luminosities around
optical clusters and test the  reliability of our X-ray luminosity measurement
using existing known X-ray clusters.  Basic X-ray properties of these clusters
are investigated  in Section  \ref{sec_correlation}.  Finally, we  present our
conclusions in Section \ref{sec_summ}.

Throughout this paper, we use  the $\Lambda$CDM cosmology whose parameters are
consistent  with the  7-year data  release of  the WMAP  mission: $\Omega_{\rm
  m}=0.275$,  $\Omega_{\rm \Lambda}=0.725$,  $h=0.702$,  and $\sigma_8=0.816$,
where the reduced Hubble constant, $h$, is defined through the Hubble constant
as  $H_0=100h~{\rm  km~s^{-1}~Mpc^{-1}}$  (Komatsu  et  al.   2011).   If  not
specified otherwise, we  use $\Lx$, $\Fx$, $\Tx$, $M_h$,  $r_{200}$, to denote
the X-ray luminosity, flux, gas temperature, halo mass and halo radius of each
X-ray  cluster.   These  quantities  are  quoted in  units  of  $10^{44}  {\rm
  erg~s^{-1}}$, $10^{-12} {\rm erg~s^{-1}~cm^{-2}}$, ${\rm keV}$, $\msunh$ and
$\mpch$, respectively.

\begin{figure*}
\center{\includegraphics[height=8.0cm,width=15.0cm,angle=0]{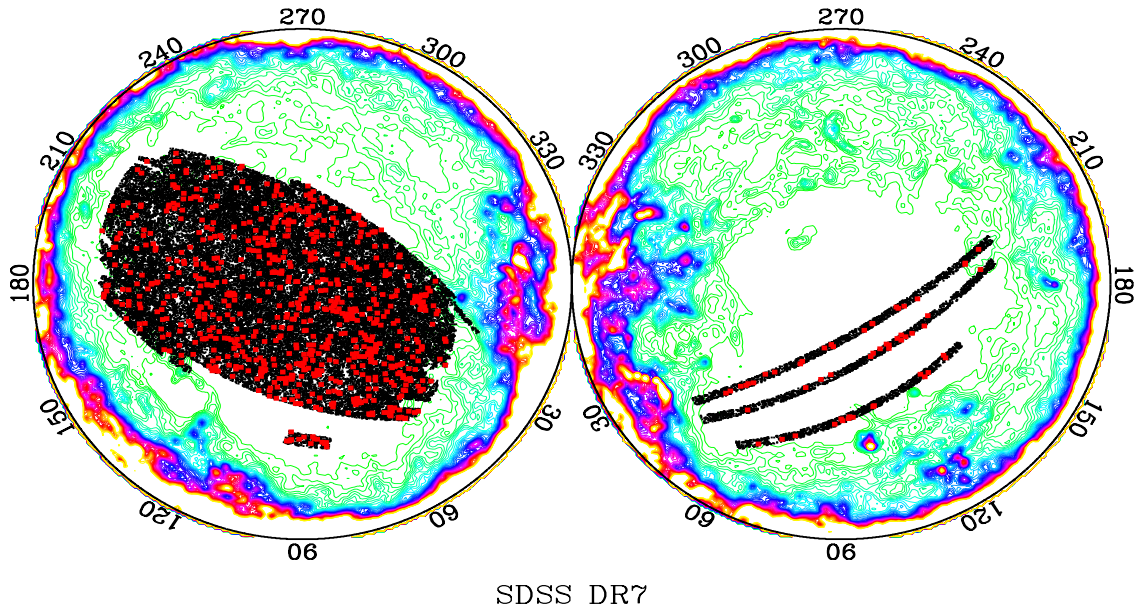}}
\caption{The  sky   coverage  of  the   SDSS  DR7  clusters  with   mass  $\ga
  10^{13}\msunh$ (black area), overlaid  on the galactic extinction contours
  of  Schlegel,  Finkbeiner  \&  Davis  (1998).   The  red  squares  show  the
  distribution of X-ray clusters with $S/N>3$ detections.}
\label{fig:skycover}
\end{figure*}

\section{The SDSS DR7 Galaxy and Group catalogs}
\label{sec_data}

The optical  data used  in our analysis  is taken  from the SDSS  galaxy group
catalogs of Yang  \etal (2007; hereafter Y07), constructed  using the adaptive
halo-based group finder  of Yang \etal (2005), here updated  to Data Release 7
(DR7).   The parent  galaxy catalog  is  the New  York University  Value-Added
Galaxy catalog (NYU-VAGC; Blanton \etal 2005) based on the SDSS DR7 (Abazajian
\etal  2009), which  contains  an independent  set  of significantly  improved
reductions.  DR7  marks the completion of  the survey phase  known as SDSS-II.
It  features  a  spectroscopic  sample  that  is now  complete  over  a  large
contiguous  area of  the  Northern Galactic  cap,  closing the  gap which  was
present in previous data releases.   From the NYU-VAGC, we select all galaxies
in  the Main  Galaxy Sample  with an  extinction-corrected  apparent magnitude
brighter than $r=17.72$,  with redshifts in the range $0.01  \leq z \leq 0.20$
and  with a  redshift completeness  ${\cal C}_z  > 0.7$.   The  resulting SDSS
galaxy catalog contains a total of  $639,359$ galaxies, with a sky coverage of
7748 square  degrees.  Note  that a  very small fraction  of galaxies  in this
catalog  have redshifts  taken from  the  Korea Institute  for Advanced  Study
(KIAS)    Value-Added    Galaxy    Catalog    (VAGC)   (e.g.     Choi    \etal
2010)\footnote{These were kindly provided to us by Yun-Young Choi and Changbom
  Park.}.  There are $36,759$ galaxies  that do not have redshift measurements
due  to fiber  collisions, but  are assigned  the redshifts  of  their nearest
neighbors.

In  this  study,  in  order  not  to miss  any  potential  group  members  for
cross-identification, we  use the group  catalog which is constructed  for all
the galaxies, where model magnitudes are used for the group finding. In total,
there are  $472,416$ groups  in our catalog  within which about  $23,700$ have
three member galaxies or more.  Following  Y07, for each group in the catalog,
we  estimate   the  corresponding   halo  mass  using   the  ranking   of  its
characteristic stellar  mass, defined as the  total stellar mass  of all group
members  with $\rmag \leq  -19.5$.  Here  the halo  mass function  obtained by
Tinker et  al.  (2008)  for WMAP7  cosmology and $\Delta=200$  is used  in our
calculation,  where $\Delta$  is  the  average mass  density  contrast in  the
spherical halo.  We  indicate the group mass obtained this  way by $M_{\rm h}$
or  $M_{200}$. Note that  groups whose  member galaxies  are all  fainter than
$\rmag =  -19.5$ cannot be assigned a  halo mass with this  method.  For these
systems, one  could in principle  use the relation  between halo mass  and the
stellar mass of  the central galaxy obtained by Yang  \etal (2012) to estimate
their halo masses.  However, since our  main focus is on probing the X-ray and
optical properties of  massive clusters, we do not need  halo masses for these
low mass groups.

According to our  definition of halo mass $M_{\rm h}$  ($M_{200}$), a halo has
an average  overdensity of 200 times  the mean density of  the universe within
its `virial radius', $r_{200}$, which is given by
\begin{equation}\label{eq:r200}
r_{\rm 200}=\left[\frac{M_{200}}{\frac{4\pi}{3}\times200\Omega_{\rm m}\times
\frac{3H_0^2}{8\pi G}} \right]^{1/3} (1+z)^{-1}\,,
\end{equation}
where $z$  is the  redshift of the  group (i.e.,  the average redshift  of its
members).  Tests with  detailed mock galaxy redshift surveys,  which take into
account various  survey selection effects,  uncertainties in the  group finder
and halo  mass estimations, have shown  that the statistical  error in $M_{\rm
  h}$ is of the  order of 0.3 dex and quite independent  to halo mass (see Y07
for details).

From the  above SDSS  DR7 group catalog,  we select 64,646  candidate clusters
with masses $\ga 10^{13}\msunh$, which  serves as our input cluster sample for
X-ray detection in the RASS image data.  As an illustration, the black dots in
Fig. \ref{fig:skycover}  show the sky coverage  of these clusters  in the SDSS
DR7.

\begin{figure*}
\center{\includegraphics[height=7.0cm,width=15.0cm,angle=0]{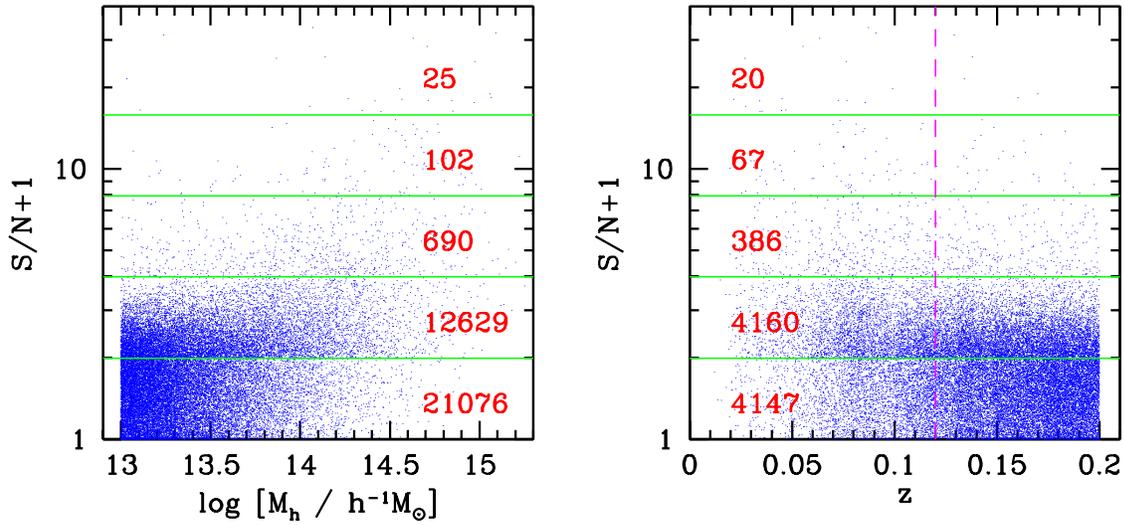}}
\caption{The $S/N$ ratios  of X-ray clusters as a function  of halo mass (left
  panel) and redshift (right panel).  The number of X-ray clusters within each
  $S/N$ ratio  bin (divided by the  horizontal lines) is  marked for reference
  (for all clusters in the left panel and for clusters with $z\le 0.12$ in the
  right panel). }
\label{fig:SN}
\end{figure*}


\section{Measuring the X-ray luminosities of clusters from RASS
  imaging data}
\label{sec_OTX}


The main goal of this paper is to obtain the X-ray luminosities/signals around
known optical clusters in the SDSS DR7 so that the resulting X-ray information
can  be used together  with the  optical information.   In addition,  this may
enable  us  to obtain  a  larger X-ray  cluster  catalog  with reliable  X-ray
detections from  the RASS (see  Wang et al.   2011 for a  cross-identified 201
entries in the SDSS DR7 regions).

\subsection{The X-ray luminosities around optical clusters}
\label{sec:X-ray}

The algorithm we use to measure the X-ray luminosity is the same as in Shen at
al.  (2008), which is a modified version of the growth curve cluster detection
method of  B\"ohringer et  al.  (2000; 2004),  and is  referred to as  the OTX
(Optical to  X-ray) code  in the following.   In our OTX  detection algorithm,
known information  (e.g.  $r_{200}$, $\sigma$,  and the positions of  the most
massive galaxies) from the optical cluster catalog are extensively used. Here,
we outline the main steps in the OTX code.

{\bf Step 1:} Starting from a  given optical cluster (with $\log M_{\rm h} \ga
13$), we  sort the  stellar masses of  its member  galaxies and find  the most
massive galaxies (MMGs).   If the cluster has more than  4 member galaxies, we
keep only 4 MMGs.

{\bf Step 2:} For  each cluster, we locate the RASS fields  where its MMGs (up
to 4)  reside (Voges et al.  1999).   We then apply a  maximum likelihood (ML)
detection algorithm  on the RASS image of  each of the fields  and generate an
X-ray source catalog which includes sources with detection likelihood $L>7$ in
the 0.5-2.0 keV  band (see Voges et al. 1999 for  details).  By matching X-ray
sources within  $0.3r_{200}$ from the MMGs,  we determine the  X-ray center of
the cluster  using the maximum X-ray  emission density point (see  Shen et al.
2008 for details). For those  clusters without any X-ray emission density
  points  that   have  likelihood  $L>7$   detections,  we  use  the   MMG  in
  consideration as the X-ray center.

{\bf Step 3:} We mask out the ML detected sources that are not associated with
the  cluster.  For example,  the QSOs  and stars  cross-matched from  RASS and
SDSS-DR7 ($\sim  9200$) within $r_{200}$  from the X-ray center  defined above
are masked  out (see Shen  et al.  2008  for details).  We then  determine the
X-ray background  from an annulus centered  on the X-ray center  with an inner
radius of $r_{200}$ and width  of 6 arcmin.  After subtracting the background,
we generate  a 1-dimensional source  count rate profile and  the corresponding
cumulative  source count  rate  profile  as functions  of  radius.  The  X-ray
extension radius  is set as $r_X=0.5r_{200}$ (adopting  smaller radius results
in too few pixels for low-mass groups at high redshift).  As we have tested by
changing  this  value from  $r_X=0.5r_{200}$  to  $r_X=r_{200}$,  none of  our
results are significantly impacted. Here  we do not use the traditional (blind
search)  growth curve  method in  determining  the $r_X$  (see B\"ohringer  et
al. 2000 for detail), as such a  method may lead to bias due to the relatively
high background in RASS.

{\bf Step 4:}  Calculate the X-ray luminosity $\Lx$.   We integrate the source
count rate  profile inside $r_X$  and get the  total source count rate  of the
cluster.    We  then   assume  that   the   X-ray  emission   has  a   thermal
spectrum\footnote{We  use  the halo  mass  provided  in  Y07 to  evaluate  the
  velocity dispersion of  the cluster, so as to make an  estimate of the X-ray
  cluster gas  temperature through $T=(\sigma /403 \rm  {km s^{-1}})^2$ (White
  et al.  1997,  Shen et al. 2008).   } with a temperature $T$,  and gas metal
abundance of one third of the solar value to make a conversion from the source
count  rates to  X-ray  fluxes (and  to  X-ray luminosities  according to  the
cosmology used in  this paper).  Finally, we make  a $\beta$ profile extension
correction to make up the X-ray luminosity missed in the range $r_X<r<r_{200}$
(see B\"ohringer  et al. 2000; 2004;  Shen et al.   2008).  For each of  the 4
MMGs, we calculate the related $\Lx$ following steps 2-4.

{\bf  Step 5:}  Determine the  central galaxy.   We identify  the central
  galaxy of  the X-ray cluster from  the 4 MMGs.   If the values of  $\Lx$ are
  different for  the 4 MMGs,  the central  is defined to  be the one  that has
  $S/N>1.0$ and with  the maximum $\Lx$. If more than  one MMGs have $S/N>1.0$
  and the  difference of their $\Lx$ is  less than the minimum  of their $\Lx$
  errors, we select  the one that has the  maximum $M_{\ast}^{1/3}/D_p$ as the
  central galaxy,  where $M_{\ast}$ is the  stellar mass of  the candidate and
  $D_p$  is  the  projected  distance  between the  candidate  and  the  X-ray
  center. The factor $M_{\ast}^{1/3}/D_p$  used here is somewhat arbitrary but
  is a  balance between the mass of  the candidate galaxy and  its distance to
  the X-ray center. We have tested that  changing the power from $1/3$ to 0 or
  1 yields results that are not significantly different. If none of the 4 MMGs
  has $S/N>1.0$,  we assign the  first MMGs as  the central. Once  the central
  galaxy is determined, we remove other MMGs and their associated measurements
  from our  X-ray cluster list.   Thus defined (chosen), the  central galaxies
  are in general associated with the X-ray flux centriods.

We run  our OTX  code to search  for X-ray  signals around all  64,646 optical
clusters with mass $\log M_{\rm h} \ga 13$. A total of 34,522 ($\sim 53\%$) of
these clusters  have a signal-to-noise ratio  $S/N > 0$  after subtracting the
background signal. Here the signal-to-noise ratio on the X-ray cluster flux is
calculated using
\begin{equation}\label{eq:SN}
{S/N} = \frac{R_{S}\, T}{\sqrt{R_{S}\, T + R_{B}\, T }},
\end{equation}
(e.g.  Henry  et al.   2006), where $R_{S}$  is the  source count rate  in the
aperture of radius  $r_{200}$, $R_{B}$ is the background  counting rate scaled
to the  aperture of  radius $r_{200}$,  and $T$ is  the exposure  time.  These
signal  to noise  ratios  suffer  from two  problems:  $(i)$ source  confusion
(projection effect):  more than one  cluster contribute to the  X-ray emission
within $0.5r_{200}$, causing  the X-ray flux to be  over-estimated, and $(ii)$
low $S/N$,  especially for high redshift  and low halo mass  sources.  To make
maximal  use of the  X-ray information,  we apply  the following  algorithm to
address these two problems.

First for source  confusion, quite a large fraction  of X-ray clusters in
  our sample suffers from this projection effect. In order to keep the maximum
  number  of X-ray  clusters,  we  carry out  the  following X-ray  luminosity
  division  among the  projected multi  X-ray cluster  systems. (1)  For these
  systems, we make  use of the {\it average} $\Lx$ -  $M_h$ relation to obtain
  the `expected'  X-ray luminosities of  the individual X-ray  clusters.  The
average $\Lx$-$M_h$ relation is  an important relation in probing cosmological
parameters through  the X-ray  luminosity function of  clusters, and  has been
extensively  discussed and  calibrated in  the literature  (e.g.   Reiprich \&
B\"ohringer  2002; Stanek \etal  2006; Vikhlinin  \etal 2009;  Leauthaud \etal
2010;  Arnaud  \etal  2010;  Mantz  \etal  2010).   Since  we  focus  only  on
low-redshift  ($0.01\leq z\leq  0.2$)  X-ray clusters,  with rest-frame  X-ray
luminosities measured  in the broad  ROSAT passband ($0.1$-$2.4$ keV),  we use
the $\Lx$-$M_h$ relation obtained by Mantz \etal (2010).  By investigating the
properties of 238 X-ray flux-selected  galaxy clusters, Mantz \etal (2010) got
the  following  X-ray luminosity-mass  scaling  relation  which  is free  from
Malmquist and Eddington biases,
\begin{equation}\label{eq:LMz}
\log \frac{L_{500\rm  c}}{E(z)~10^{44} {\rm erg~ s^{-1}}}=0.80+1.34~\log 
\frac{E(z)~M_{500\rm  c}}{10^{15} M_{\odot}}\,,
\end{equation}
and which has an intrinsic scatter $\sigma\sim 0.185$. In this relation, $E(z)
= H(z)/H_0$,  $M_{500\rm c}$ is  the halo mass  of the X-ray cluster  within a
radius $r_{500\rm  c}$, defined  as the radius  within which the  average mass
density  is  $500$  times the  critical  mass  density  of the  Universe,  and
$L_{500\rm c}$ is the total X-ray luminosity within $r_{500\rm c}$. Piffaretti
\etal (2011) employed  an iterative algorithm to calculate  $L_{500\rm c}$ for
sources   with   available  aperture   luminosities,   and  found   $L_{500\rm
  c}/\Lx=0.91$.   With  this  transformation,  we obtain  $M_{500\rm  c}$  and
$r_{500\rm c}$ for  an X-ray cluster with given $\Lx$. The  final step is then
to convert  $M_{500\rm c}$ to $M_{200}$,  and $r_{500\rm c}$  to $r_{200}$ for
consistency with  our halo definition,  i.e.  the average overdensity  is 200.
Here we  assume that dark matter  halos follow a NFW  density profile (Navarro
\etal  1997) with  concentration  parameters given  by the  concentration-mass
relation of Maccio \etal (2007).  Based on this assumption, we have
\begin{eqnarray} \label{eq:convert}
r_{200} &\simeq& 2.70\times r_{500\rm c} \,, \nonumber \\
M_{200} &=& M_{500\rm c}\times \frac{200}{500}\times\Omega_{\rm
  m}\times \left( \frac{r_{200}}{r_{500\rm c}}\right)^3 \,.
\end{eqnarray}
Based on the above relations, we can obtain a rough estimate of the `expected'
average $\Lx$ for each cluster and the corresponding X-ray flux.  (2) For
  multi-cluster systems, we add up all the `expected' X-ray {\it fluxes} and
  obtain a contribution fraction for each cluster, `$i$', in the system,
\begin{equation}\label{eq:f_contri}
f_{mult,i} = F_{{\rm X},i} / \Sigma_i F_{{\rm X},i}\,.
\end{equation}
This parameter is then applied to each cluster in the multi-cluster system to
partly take into account the projection effect when quantities, such as $\Lx$
and $S/N$ are calculated. 

Next, we  check the  $S/N$ of  all the $34,522$  X-ray clusters  with positive
detections (i.e. positive count  rates after background subtraction). Shown in
the left and right panels of Fig.  \ref{fig:SN} are the $S/N$ distributions of
these  clusters in  halo mass  $M_h$ and  in redshift  $z$,  respectively. For
reference,  we label  the  number  of X-ray  clusters  within different  $S/N$
bins. Among  all the  X-ray clusters, $817$  have $S/N>3$, $12,629$ have $3\ge
S/N>1$, and $21,076$ have $1\ge S/N>0$.  In a blind search for X-ray clusters,
Henry  et  al.    (2006)  adopted  $S/N=4$  as  the   threshold  for  reliable
detection.  Since here  we are  performing  a counterpart  detection, we  take
$S/N=3$ as our threshold for  reliable X-ray detections, and sources with $S/N
<3$ as  tentative detections  (see Wang 2004  for a detailed  discussion about
this detection threshold).  The red squares in Fig \ref{fig:skycover} show the
projected distribution of clusters with $S/N>3$.

\subsection{Testing the performance of OTX using existing X-ray
  clusters}
\label{sec:test}

In order to test the reliability of our algorithm, we compare our measurements
with X-ray clusters currently available  in the literature.  In a recent paper
(Wang et  al. 2011),  we have performed  the cross-identification  between the
SDSS DR7  groups and known  X-ray cluster entries  obtained from a  variety of
sources:  the  ROSAT  Brightest   Cluster  Sample  (BCS)  and  their  low-flux
extensions compiled  by Ebeling  \etal (1998, 2000),  as well as  the Northern
ROSAT  All-Sky  (NORAS) and  ROSAT-ESO  Flux  Limited  X-ray (REFLEX)  samples
compiled by B\"ohringer  \etal (2000, 2004). Within the  same redshift and sky
coverage of the  SDSS DR7, we obtained an X-ray  and optically matched catalog
of 201 entries (see Appendix of Wang et al. 2011).

\begin{figure}
\center{\includegraphics[height=7.0cm,width=7.5cm,angle=0]{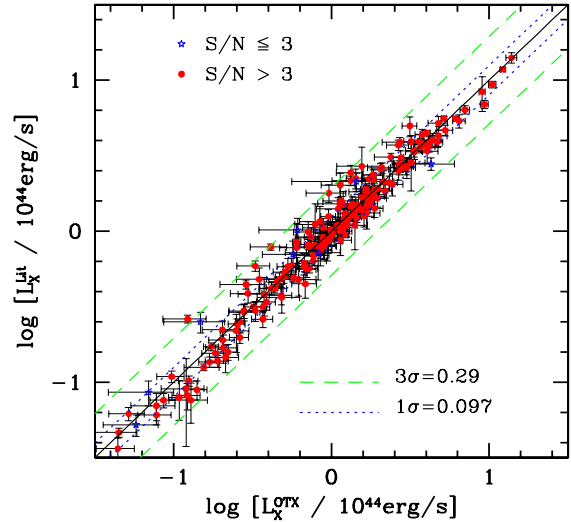}}
\caption{The X-ray luminosities of the cross-identified clusters, $\log L^{\rm
    Lit}_{\rm X}$, obtained from  literature versus $\log L^{\rm OTX}_{\rm X}$
  obtained by our OTX pipeline.}
\label{fig:X-X}
\end{figure}

Before making any further investigation, we check if the X-ray luminosities of
these clusters obtained  from our OTX code are consistent  with those given in
the  literature. We  find that  the vast  majority of  the 201  existing X-ray
clusters are  in our  X-ray cluster sample  with $S/N>3$.  Three  sources have
$S/N\le 3$ according to our OTX  algorithm, mainly due to the shallow exposure
of the RASS; in the literature they  are detected by the PSPC and/or HRI point
observations.  Note that only  13 in  our whole  sample have  $4\ge S/N  > 3$.
Given the good  cross detection, we consider our OTX  clusters with $S/N>3$ to
be  reliable.   Fig.\ref{fig:X-X}  shows  the  comparison  between  the  X-ray
luminosities obtained from literature compared to those obtained using the OTX
code. The differences between them are almost negligible, and the luminosities
obtained  in  these  two  ways  are  consistent with  each  other  within  the
1-$\sigma$ scatter  of $0.097$  dex.  Moreover, even  the three  clusters with
$S/N\le 3$ do  not show large deviations.  This test  indicates again that our
OTX code is reliable.

Within the SDSS DR7 sky coverage, our OTX pipeline extracts 817 X-ray clusters
with $S/N > 3$  with the help of optical data.  Comparing  to the existing 201
X-ray clusters, our  new X-ray cluster sample is larger by  a factor of almost
4.  Including clusters with $3>S/N \ge 1$ further increases the sample size by
another factor of $\sim 15$.

\begin{figure*}
\center{\includegraphics[height=14.0cm,width=15.0cm,angle=0]{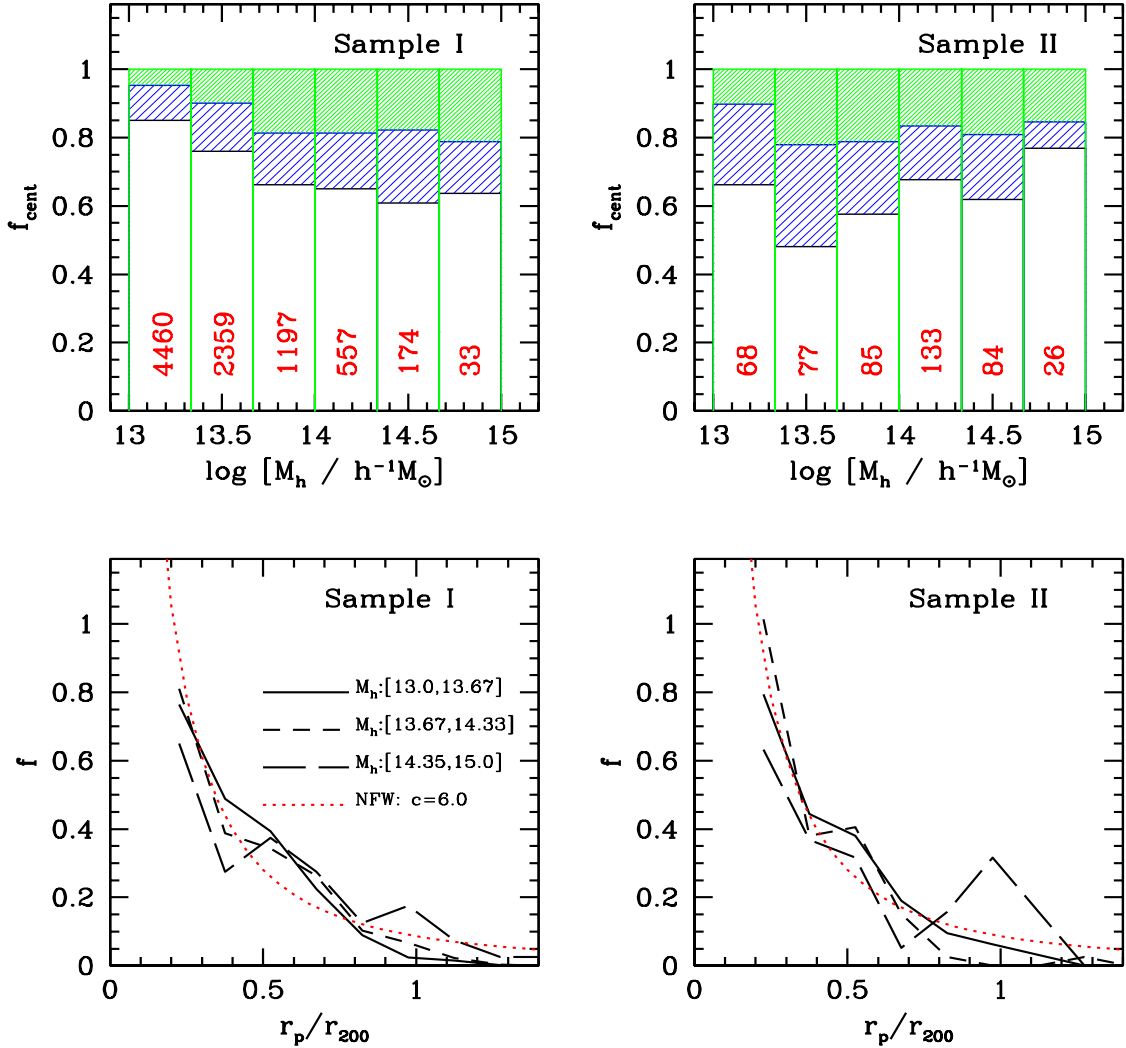}}
\caption{Shown in  the upper panels  are the accumulative fraction  of central
  galaxies that are  the MMGs, the second MMGs and others.  Shown in the lower
  panels are the distributions of the projected distances between the MMGs and
  the central  galaxies that are not  MMGs. Here results  are shown separately
  for samples {\bf I} and {\bf II} in the left and right panels, respectively.
}
\label{fig:f_cent}
\end{figure*}

\begin{figure*}
\center{ \includegraphics[height=14.0cm,width=15.0cm,angle=0]{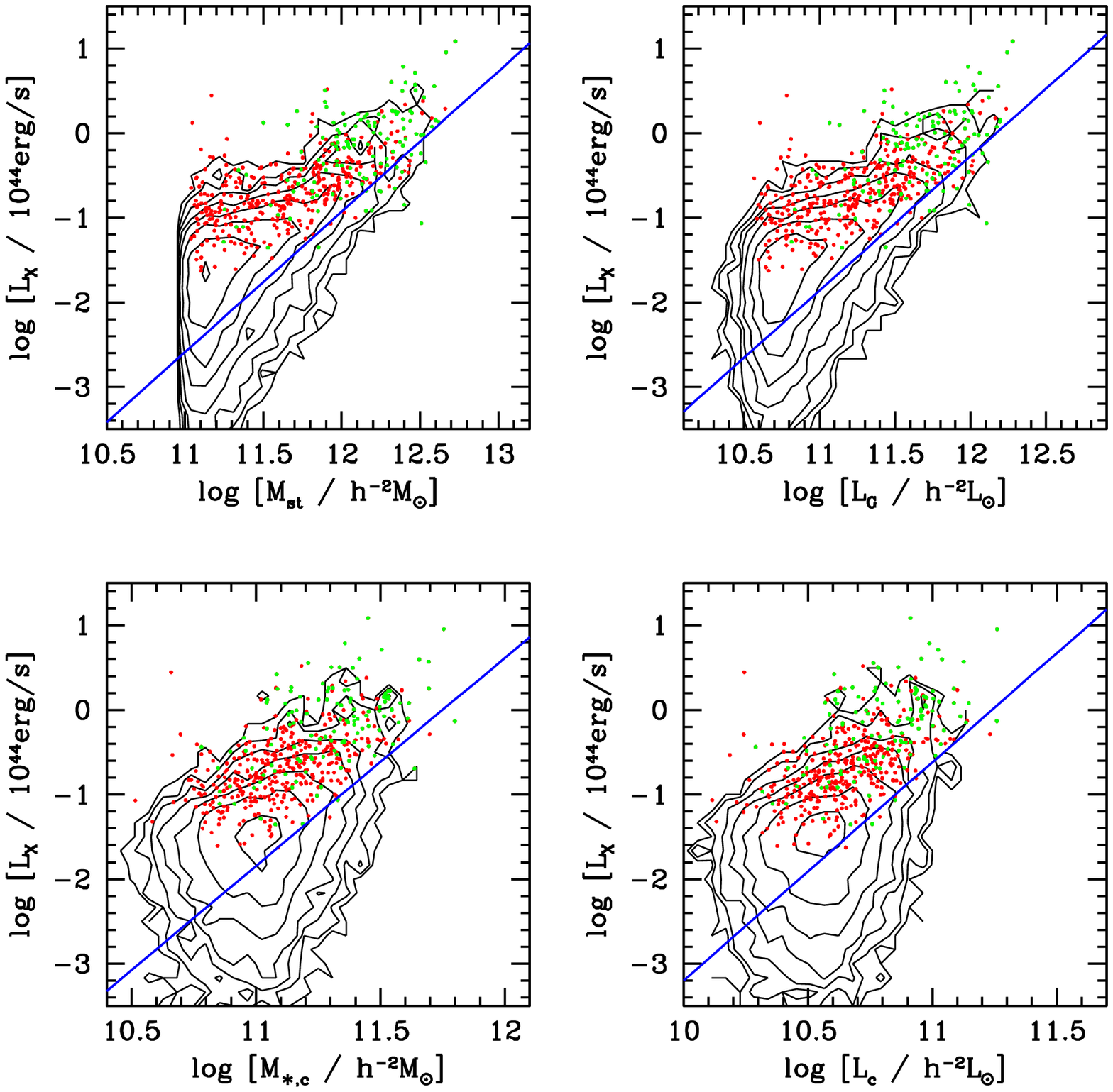}}
\caption{The  correlation of X-ray  cluster luminosity,  $\log \Lx$,  with the
  characteristic stellar  mass (upper-left  panel) and the  cluster luminosity
  (upper-right  panel), the  stellar  mass (lower-left  panel) and  luminosity
  (lower-right panel) of  the MMG.  Here results for sample  {\bf I} are shown
  as the black contours, while red  dots are results for sample {\bf II}.  For
  comparison,  we also  show  as green  dots  the results  for  the 131  X-ray
  clusters that are  obtained from the literature with  $z\le 0.12$. The solid
  line in each panel is the best fit scaling relation (see text for details).}
\label{fig:optic-Lx}
\end{figure*}

\section{General properties of our X-ray clusters}
\label{sec_correlation}

Now that we  have measured the X-ray luminosities for  the optical clusters in
SDSS  DR7,  we  can proceed  to  examine  various  properties of  these  X-ray
clusters.   Because of  observational limits  both  in X-ray  and in  optical,
properties of  clusters at higher redshift  are expected to  be less accurate.
In  particular the  group catalog  of Y07  was constructed  based  on galaxies
brighter than  $\rmag=-19.5$, corresponding to  a redshift $z\sim 0.1$  at the
magnitude limit  of the SDSS redshift  survey. As a  compromise between sample
size and reliability, we only use clusters at $z< 0.12$. Tests have shown that
this redshift  cut does not impact  any of our results,  other than increasing
the scatter  in some  of the  relations. Using the  X-ray clusters  with $z\le
0.12$ we  construct two  samples for our  investigation: sample {\bf  I} which
contains $8,780$  clusters  with $S/N>0$,  and sample {\bf II}  which contains
the  subset of $473$  clusters with  $S/N>3$. A  comparison between  these two
samples can be used to test the reliability of the cluster detection.

\subsection{Distances between the central and most massive galaxies in clusters}
\label{sec:separation}

The central galaxy in a dark matter halo plays an important role in the theory
of galaxy formation (e.g.  Mo et al.  2010) as well as in our investigation of
the   dark  matter  distribution   using  galaxy-galaxy   lensing  measurement
(e.g. Yang  et al.  2006;  Li et  al. 2013).  However,  as dark matter  is not
directly observable,  it is not  straightforward to determine which  galaxy is
the true `central' galaxy in a dark matter halo. In this paper we refer to the
galaxy that is closest to the X-ray peak as the central galaxy.  However, this
center may deviate  from that of the corresponding dark  matter halo; in fact,
in a few  percent of cases, the central galaxy thus  defined deviates from the
X-ray peak position  by up to a few arcminutes. These  offsets are provided in
our X-ray catalogue. An alternative  method for defining the `central' galaxy,
is to associate it with the MMG (e.g.  Yang et al.  2008). Once again, though,
the location of the MMG may deviate significantly from that of the dark matter
halo (e.g., Skibba et al. 2011).  In this section, we examine the distribution
of  the  separation  between  the   two  centers  identified  with  these  two
definitions.  Such  information is  useful in modelling  the dark  matter mass
distribution   around   given  clusters   with   galaxy-galaxy  weak   lensing
measurements  (e.g., Yang  et al.   2006; Johnston  et al.   2007;  Sheldon et
al.  2009;  R. Li  et  al.   2013  in preparation;  W.  Luo  et al.,  2013  in
preparation).

We first check  the fraction of central galaxies in  our X-ray cluster catalog
that are  the MMGs,  the second MMGs  and other  ranks of member  galaxies. As
shown in the upper-left panel of Fig. \ref{fig:f_cent} for sample {\bf I}, the
fraction of the MMGs that are  central galaxies change from about 65\% in very
massive clusters to about 80\% in  relative low mass clusters.  In $\sim 18\%$
of  all  the  clusters, the  central  galaxies  are  the second  most  massive
galaxies.  As  a reference, the  total number of  X-ray clusters in  each halo
mass bin is provided in the  figure.  Since the majority of our X-ray clusters
have low $S/N$, the upper-right  panel of Fig.~\ref{fig:f_cent} shows the same
distribution, but  for the X-ray clusters  with $S/N>3$ (sample  {\bf II}). In
this case,  the fraction of central galaxies that  are MMGs ranges  from $\sim
50\%$ to  $\sim 75\%$, with an average at $\sim 65\%$. 

Next, we  calculate the distances between  the `centrals' and the  MMGs in our
X-ray cluster sample. Note that here we only show results for central galaxies
that are not the MMGs; for the majority cases, the distances are zero.  In the
lower-left panel  of Fig.  \ref{fig:f_cent},  we show the distribution  of the
projected  distance, $r_{\rm  p}$, obtained  from  all the  X-ray clusters  in
sample {\bf  I}. Results are shown  separately for clusters  in different mass
bins, as indicated,  and the distance $r_{\rm p}$ is  normalized by the virial
radius of each cluster in question.  For comparison, the dotted line shows the
distribution expected for a  projected NFW profile with concentration $c=6.0$.
As one can see,  the MMGs that are not central galaxies  follows roughly a NFW
profile. Although in general we are unable to determine whether the X-ray peak
or the MMG traces the halo center better, the MMG fraction of centrals and the
central-MMG distance  distribution presented above  can help us  in separating
weak lensing signal of centrals from that of satellites in a statistical sense
(e.g., Johnston et al.  2007; W. Luo et al., 2013 in preparation; R. Li et al.
2013 in preparation).

\begin{figure}
\center{ \includegraphics[height=7.0cm,width=8.0cm,angle=0]{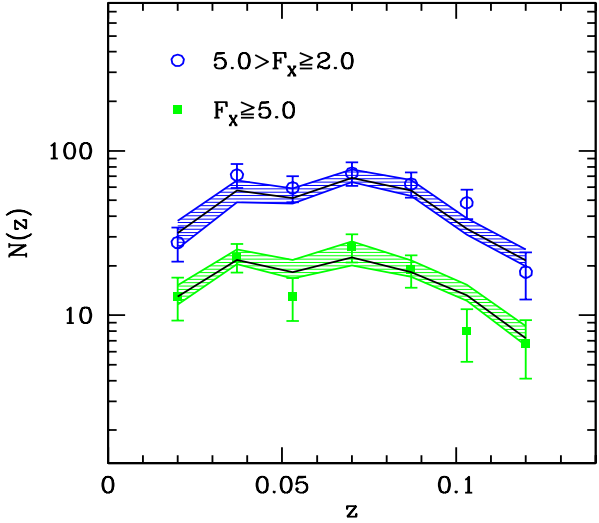}}
\caption{The  redshift distributions  of X-ray  clusters within  two different
  X-ray flux bins, as indicated using different symbols. For clarity, data for
  the low flux bin is shifted upward by a factor of 2. Error bars are obtained
  from 200 bootstrap  resamplings of the clusters. The  solid lines and shaded
  regions indicate the  corresponding best fit model predictions  and the 68\%
  confidence levels. }
\label{fig:Nz}
\end{figure}

\subsection{X-ray - optical scaling relations}
\label{sec:general}

In this subsection, we examine  the correlations of the X-ray luminosity $\Lx$
with  a  number  of other  properties  obtained  from  the optical  data:  the
characteristic stellar  mass $M_{\rm  st}$ and luminosity  $L_{\rm G}$  of the
cluster, the stellar mass and luminosity  of the MMG, and the halo mass.  

The  upper  panels of  Fig.~\ref{fig:optic-Lx}  show  the correlation  between
cluster X-ray  luminosity and characteristic  stellar mass $M_{\rm  st}$ (left
panel)  and luminosity  $L_{\rm  G}$  (right panel).  Here  contours show  the
results for  sample {\bf I} and  dots for sample  {\bf II}.  There is  a clear
trend that clusters with larger characteristic stellar masses and luminosities
are more X-ray luminous.  For a given characteristic stellar mass $M_{\rm st}$
or luminosity $L_{\rm  G}$, the typical scatter in $\log  \Lx$ is quite large:
$\sim 0.40$ for sample {\bf II}.  To  make sure that the scatter is not due to
our OTX  code, we show as  green dots the  results for the 131  X-ray clusters
selected  from  the literature  (Wang  et al.   2011)  applied  with the  same
redshift  cut $z\le  0.12$.  The fact  that  these 131  clusters show  similar
scatter suggests that this kind of scatter is likely intrinsic.

To  obtain a  rough  guide to  the  relations between  the  X-ray and  optical
properties, we fit the observational results with
\begin{equation}\label{eq:scale1}
\log \Lx =a~ (\log Y_{\rm opt} - b)\,,
\end{equation}
where  $Y_{\rm opt}=M_{\rm st}$  for characteristic  stellar mass  and $Y_{\rm
  opt}=L_{\rm G}$ for luminosity.   In general, with reliable X-ray luminosity
measurements for all  the optical clusters, so that the  sample is complete in
$M_{\rm st}$ and $L_{\rm G}$, one  could obtain directly a scaling relation as
described by  the above equation by  fitting it into  data.  Unfortunately, in
our X-ray sample,  only about 21\% of the objects  have $S/N>1.0$.  As pointed
out in Stanek et al. (2006),  using only a small fraction of reliably detected
X-ray clusters (i.e. with high $S/N$)  to obtain the scaling relations tend to
systematically  overestimate $\Lx$  for a  given optical  property due  to the
large intrinsic scatter in such a relation.  This is caused by the combination
of the  Malmquist bias due to the  fact that only X-ray  luminous clusters are
observed, and the Eddingtion bias owing to the fact X-ray fainter clusters are
more abundant  than X-ray brighter ones  (see, e.g., Wang 2004;  Stanek et al.
2006; Mantz et  al.  2010). To alleviate the impact of  such biases, we follow
the method  of Stanek et al. (2006)  to constrain the two  free parameters $a$
and $b$. This  method uses the abundance of  reliably detected X-ray clusters.
To do this we  obtain the number of clusters as a  function of redshift in two
X-ray flux bins.   The results are shown in  Fig. \ref{fig:Nz} using different
symbols.   Our test  shows that  the  vast majority  ($\ga 95\%$)  of all  the
clusters  with  $\Fx \ge  2.0\times  10^{-12}  {\rm erg~s^{-1}~cm^{-2}}$  have
$S/N>1.0$.

We constrain the  model parameters, $a$ and $b$, as  follows. Starting from an
initial guess of values of $a$ and $b$, we predict the median X-ray luminosity
for each cluster  using Eq. \ref{eq:scale1} from its  $M_{\rm st}$ (or $L_{\rm
  G}$).  Since the redshift range covered  by the clusters is small, we ignore
any possible evolution  in the scaling relations.  A  log-normal dispersion is
applied  to  the  median  X-ray  luminosity.   As pointed  out  in  Stanek  et
al. (2006),  the dispersion  itself are not  well constrained by  the redshift
distribution.  We  thus fix the dispersion, $\sigma_L=0.4$,  according to that
obtained directly from Sample {\bf II}.  We have tested that any change in the
dispersion  at the level  of $\pm  50\%$ does  not change  any of  our results
significantly.  The X-ray luminosities of {\it all} the clusters with redshift
$z\le 0.12$ obtained this way are  converted into X-ray fluxes in the observed
band taking  into account  the luminosity distances  and negative  average $K$
corrections based on the redshifts  of individual clusters (B\"ohringer et al.
2004).   A mock  `X-ray cluster  catalogue' is  then constructed,  and  we can
calculate the redshift  distributions of the mock clusters  in different X-ray
flux ranges, as shown in Fig.  \ref{fig:Nz}.  These redshift distributions are
then  compared  with those  obtained  directly  from  the observed  sample  to
constrain the scaling relations.  In  practice we define a goodness-of-fit for
each model using
\begin{equation}
\label{chisq}
\chi^2 = \sum \left[ {N(z) - \hat{N}(z) \over \Delta \hat{N}(z)} \right]^2\,,
\end{equation}
where $\hat{N}$ and $\Delta\hat{N}$ are the observed average number and error,
respectively. To  explore the  best fit  values and the  freedom of  the model
parameters, we follow Yan, Madgwick \&  White (2003; see also van den Bosch et
al.  2005) and use a Monte-Carlo  Markov Chain (hereafter MCMC) to explore the
parameter  space.   We start  our  MCMC  from an  initial  guess  and allow  a
`burn-in'  of 1000  random walk  steps  for the  chain to  equilibrate in  the
parameter space. At  each step in the  chain we generate a new  trial model by
drawing  the   shifts  in  the  free  parameters   from  independent  Gaussian
distributions.  The probability of accepting the trial model is assumed to be
\begin{equation}
\label{probaccept}
P_{\rm accept} = \left\{ \begin{array}{ll}
1.0 & \mbox{if $\chi^2_{\rm new} < \chi^2_{\rm old}$} \\
{\rm exp}[-(\chi^2_{\rm new}-\chi^2_{\rm old})/2] & \mbox{if
$\chi^2_{\rm new} \geq \chi^2_{\rm old}$} \end{array} \right.
\end{equation}
with $\chi^2$ given by eq.~(\ref{chisq}).

We construct a MCMC of $1$ million steps, with an average acceptance rate of
$\sim 25\%$. We have tested its convergence using the 'convergence ratio' $r$
as defined in Dunkley et al.  (2005).  In all cases $r<0.01$ is achieved for
each parameter.  To suppress the correlation between neighboring steps in the
chain, we thin the chain by a factor of $100$.  This results in a final MCMC
consisting of $10,000$ independent models that sample the full posterior
distribution.  Among these models, we obtain the $68\%$ range of the best
parameters with smaller $\chi^2$, and the best fit values are those with the
smallest $\chi^2$ value.  The resulting scaling relations for $M_{\rm st}$ and
$L_{\rm G}$ are:
\begin{eqnarray}\label{eq-scaling}
\log \Lx &=& 1.66^{+0.06}_{-0.17}~(\log M_{\rm st}-12.56^{+0.08}_{-0.03})\,, \nonumber \\
\log \Lx &=& 1.59^{+0.16}_{-0.08}~(\log L_{\rm G}  - 12.17^{+0.04}_{-0.07})\,,
\end{eqnarray}
respectively,  each  of  the quantities  having  the  same  units as  in  Fig.
\ref{fig:optic-Lx}.   Here the  superscript  and subscript  of each  parameter
indicate the 68\% confidence level.   The best fit scaling relations are shown
as the  solid lines  in the corresponding  panels of  Fig. \ref{fig:optic-Lx}.
The scaling  relations so  obtained are steeper  and smaller than  those given
directly by  the small set  of 473 X-ray  clusters with $S/N>3$  which suffers
from Malmquist and Eddingtion biases.   The best fit redshift distributions of
the X-ray  clusters are shown as  the solid lines in  Fig.~\ref{fig:Nz} in the
corresponding X-ray  flux ranges,  with the shaded  areas representing  the 68
percentiles  of the $10,000$  MCMC. Here  results are  shown only  for $M_{\rm
  st}$, and the results are very similar for other optical quantities.

Next  we investigate  the correlation  between  the X-ray  luminosity and  the
stellar mass $M_{\ast, c}$ (luminosity $L_{\rm c}$) of the central galaxy of a
cluster. Here  we refer to the  most massive or brightest  cluster galaxies as
the central galaxies.  The bottom row panels of Fig.  ~\ref{fig:optic-Lx} show
these  correlations,  with $\Lx$-$M_{\ast,  c}$  in  the  left-hand panel  and
$\Lx$-$L_{\rm c}$ in  the right-hand panel.  It is  obvious that clusters with
brighter X-ray  luminosities on  average have central  galaxies that  are more
massive and more luminous.  The scatter in $\log \Lx$ is similar to that shown
in the upper panels for the characteristic stellar mass and luminosity.  Using
the   same   method  as   for   $M_{\rm  st}$   and   $L_{\rm   G}$,  we   fit
Eq.\,(\ref{eq:scale1}) to  the redshift distribution of X-ray  clusters in the
same three flux bins as  shown Figure~\ref{fig:Nz}, and obtained the following
scaling relations of $\log \Lx$ with $\log M_{\ast, c}$ and $\log L_c$:
 \begin{eqnarray}
\log \Lx &=& 2.46^{+0.18}_{-0.15}~(\log M_{\ast, c} - 11.75^{+0.03}_{-0.03})\,, \nonumber \\ 
\log \Lx &=& 2.58^{+0.12}_{-0.11}~(\log L_c - 11.24^{+0.03}_{-0.04})\,.
\end{eqnarray}
These  relations are shown  as the  solid lines  in the  bottom row  panels of
Fig.~\ref{fig:optic-Lx}.

\begin{figure}
\center{\includegraphics[height=7.0cm,width=7.5cm,angle=0]{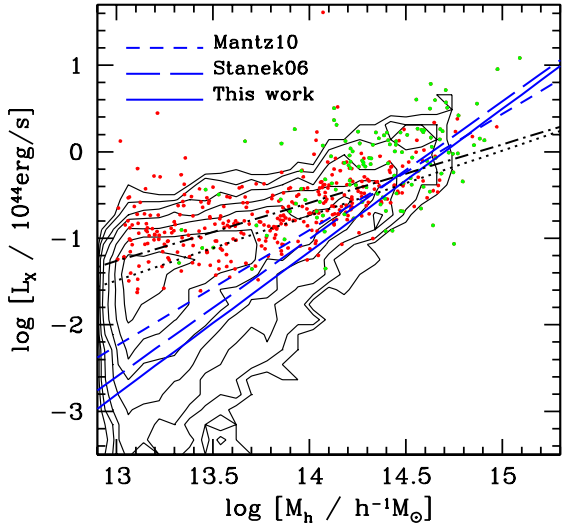}}
\caption{The X-ray cluster luminosity, $\log L_{\rm X}$, v.s.  halo mass $\log
  M_{\rm h}$.  The black contours show results for sample {\bf I}, while red
  dots for sample {\bf II}. For comparison, the green dots show the 131 X-ray
  clusters obtained from the literature with $z\le 0.12$.  The solid,
  long-dashed and dashed lines are the average $\Lx-M_h$ relations obtained in
  this work, by Stanek et al. (2006) and by Mantz \etal (2010), respectively.
  The dot-dashed and dotted lines are used to separate clusters with $S/N>3$
  and $S/N>2$ into two subsamples with similar numbers of clusters. }
\label{fig:LX-Mh}
\end{figure}

We also investigate the correlation between the X-ray luminosity $\Lx$ and the
halo  mass, and  the result  is shown  in Fig.~\ref{fig:LX-Mh}.   Although the
scatter is  large, in general $\Lx$  is positively correlated  with $M_h$.  In
order  to obtain an  unbiased scaling  relation between  $\log \Lx$  and $\log
M_h$,  we proceed  as follows.   As pointed  out in  Yang et  al.  (2007), the
typical scatter in  halo mass estimation based on the  ranking of $M_{\rm st}$
is about 0.3dex.  In order to take  such scatter into account we obtain a halo
mass for each  cluster with the following  steps : (1) Start from  a halo mass
function and extract  a list of `ture' halo masses within  the SDSS DR7 volume
at redshift  $z\le 0.12$; (2) Add  a log-normal deviation to  each `true' halo
mass with  1-$\sigma$ scatter at 0.3  and obtain a `scattered'  halo mass; (3)
Rank the  `scattered' halo masses and  associate them to the  ranks in $M_{\rm
  st}$; (4)  Assign the`true' halo  mass to each  group according to  the link
between the  true and `scattered' masses.  This approach gives  a better model
for the mass {\it distribution} of groups  but not for the halo masses of {\it
  individual} groups.  Finally,  using the similar method as  above, we fit to
the redshift distribution  of X-ray clusters, and obtain  the best fit scaling
relation between $\log \Lx$ and $\log M_h$,
\begin{equation}\label{eq:Lx-Mh}
\log \Lx = 1.65^{+0.04}_{-0.02}~(\log M_h - 14.70^{+0.01}_{-0.01}) \,,
\end{equation} 
which is shown in Fig.~\ref{fig:LX-Mh} as the solid line.  Our test shows that
using the original  assigned halo masses together with  $\sigma_L=0.4$ for the
X-ray   luminosity,    the   resulting   scaling    relation   is   consistent
Eq. \ref{eq:Lx-Mh}  at the 2-$\sigma$  level. For comparison,  the long-dashed
and dashed lines in  Fig.~\ref{fig:LX-Mh} show the average $\Lx-M_h$ relations
obtained by  Stanek et al. (2006)  and Mantz \etal  (2010), respectively.  The
scaling  relation  we get  is  in quite  good  agreement  with these  previous
studies, especially with that obtained by Stanek \etal (2006).

\begin{figure*}
\center{\includegraphics[height=14.cm,width=15cm,angle=0]{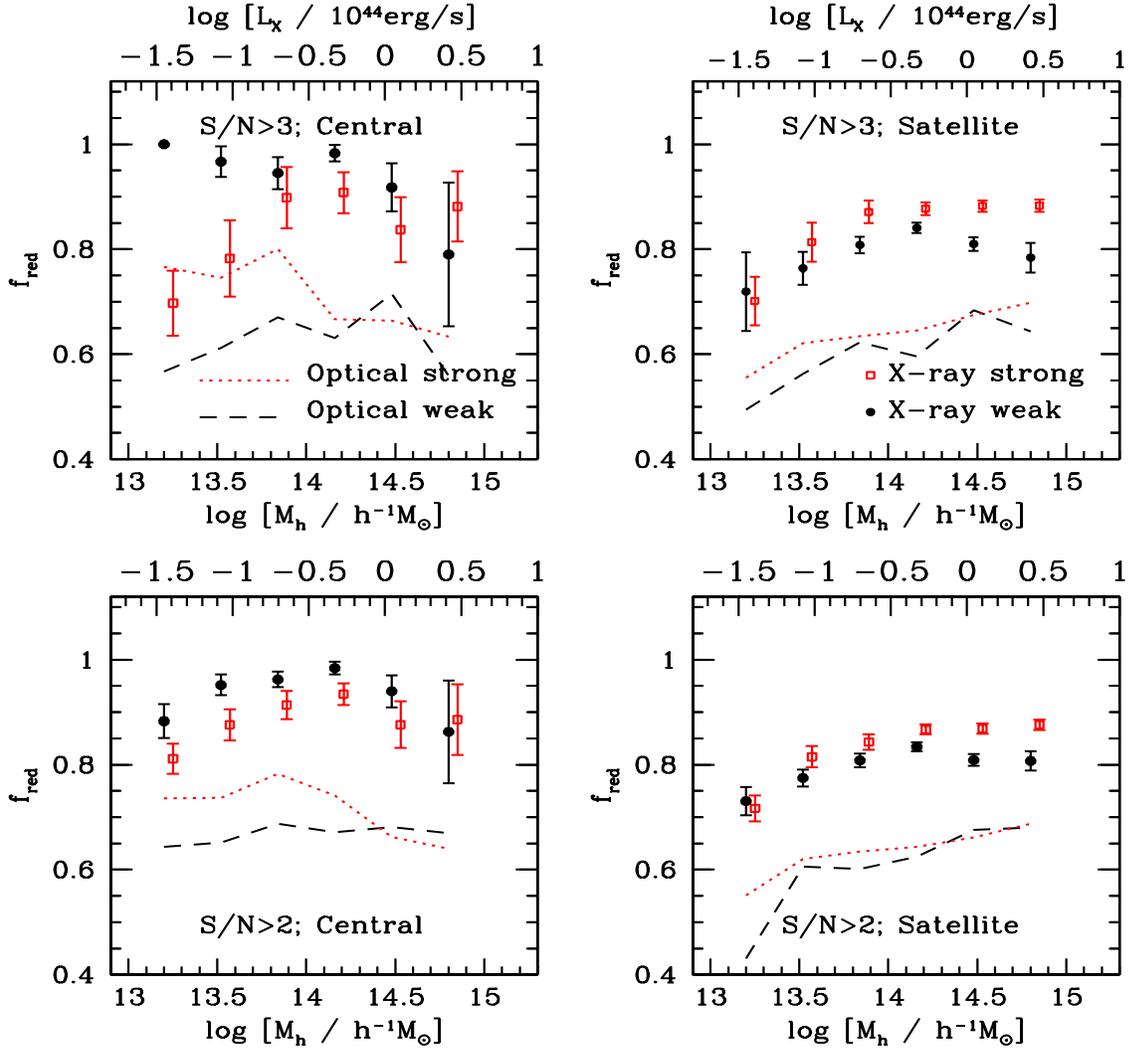}}
\caption{Upper  panels:  the red  fraction  of  central  (left) and  satellite
  (right) galaxies in clusters with  strong (squares) and weak (circles) X-ray
  emissions in Sample  {\bf II} (with $S/N>3$).  Lower  panels: similar to the
  upper panels  but for  clusters with $S/N>2$.   The error bars  are obtained
  from  200  bootstrap resamplings  of  all  the  clusters in  question.   For
  comparison, the dotted  and dashed lines are results  for `optically strong'
  and `optically  weak' clusters, respectively.  The related  average $\Lx$ in
  each sample can be  read from the top label of each  panel. And for clarity,
  the red fraction is shifted downwards by a factor of 0.2.}
\label{fig:red_v}
\end{figure*}

\begin{figure*}
\center{\includegraphics[height=14.0cm,width=15.0cm,angle=0.0]{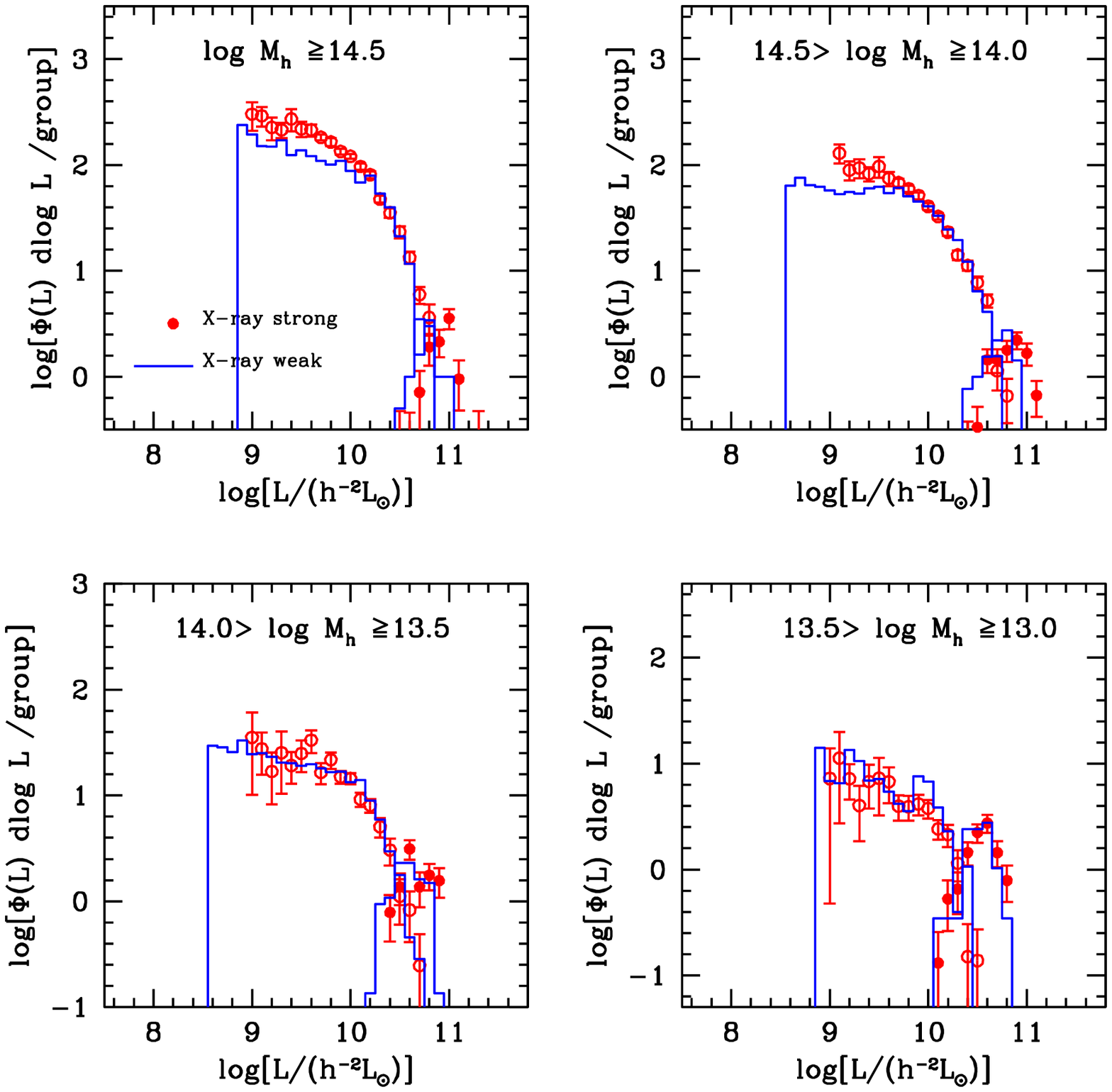}}
\caption{The  conditional   luminosity  functions  of   clusters  with  strong
  (circles)  and  weak  (histograms)  X-ray emissions.   The  contribution  of
  central and satellite  galaxies are plotted separately.  The  error bars are
  obtained from 200 bootstrap resampling of all the clusters in question. }
\label{fig:CLF}
\end{figure*}

\subsection{Galaxy properties in clusters of different X-ray luminosities}
\label{sec_diff}

As  demonstrated  above,  for  given  optical  properties  the  cluster  X-ray
luminosities  $\Lx$ have  large  variations (see  also  Castander \etal  1994;
Lubin,  Mulchaey \&  Postman  2004;  Stanek \etal  2006;  Popesso \etal  2007;
Castellano  et al.   2011;  Balogh et  al.   2011).  The  question is  whether
clusters of the  same mass (or similar optical  properties) but with different
X-ray contents  contain different galaxy  populations.  To shed light  on this
question, we  divide our 473 X-ray clusters  into 2 subsamples in  the $\Lx$ -
$M_h$ space in such a way that each subsample contains about half of the total
number at a given $M_h$.  The separation line is shown in Fig.~\ref{fig:LX-Mh}
as the dot-dashed  line.  We refer clusters above and  below the dividing line
as ``X-ray  strong" and ``X-ray  weak", respectively.  We examine  whether the
galaxy populations  in these  two subsamples are  different.  We  have checked
various  optical properties  as  a function  of  cluster mass  for both  these
subsamples, including the stellar  mass, $r$-band luminosity and concentration
of the  central galaxy, the  stellar mass, velocity dispersion  and luminosity
gap (the magnitude difference between the first and second brightest galaxies)
of member  galaxies.  None  of these reveal  any indication for  a significant
difference between X-ray strong and weak clusters.

An  exception is the  red fractions  of central  and satellite  galaxies.  The
upper panels of Fig.~\ref{fig:red_v} show  the red fractions of central (left)
and satellite (right) galaxies as a  function of cluster mass, $M_h$, both for
clusters with strong (open squares)  and weak (open circles) X-ray emission in
Sample {\bf II}.   Here galaxies have been split in  red and blue populations,
using the following magnitude-dependent color criterion:
\begin{equation}\label{quadfit}
^{0.1}(g-r) = 9.97 -0.0651 \, x - 0.00311\,x^2\,,
\end{equation}
where $x=\rmag  + 23.0$, which is similar  to that used in  Yang \etal (2008).
Clusters  with masses above  and below  $\sim 10^{14}\msunh$  reveal different
behaviors.  For $M_h \ga 10^{14}\msunh$, clusters with stronger X-ray emission
have  higher red  satellite fraction:  $\sim  90\%$ in  X-ray strong  clusters
versus  $\sim 80\%$ in  X-ray weak  clusters. This  result is  in quantitative
agreement with the findings of Popesso et al.  (2007) based on a significantly
smaller sample of X-ray clusters.   For clusters with $M_h \la 10^{14}\msunh$,
the trend disappears in satellite  galaxies but appears in central galaxies in
an  opposite direction.   For $M_h  \sim 10^{13}\msunh$,  the red  fraction of
central galaxies in X-ray strong clusters  is about 70\%, much lower than that
in X-ray weak clusters, which is  $\sim 95\%$. To test the robustness of these
results, we use a larger sample by adding clusters with lower $S/N$. The lower
panels of Fig.~\ref{fig:red_v}  show the same results as  the upper panels but
for a  total of 1193 clusters with $S/N>2$.   Here the X-ray strong  and X-ray
weak   clusters    are   separated   using   the   dotted    line   shown   in
Fig.\ref{fig:LX-Mh},  which  splits  the  sample in  subsamples  that  contain
similar numbers of  clusters at a given halo mass.  Clearly, the general trend
here is similar to that obtained using the smaller, higher-significance sample
of clusters with  $S/N>3$.

These  results may  indicate that  there is  a transition  in the  gas heating
mechanism at  a halo mass $\sim  10^{14}\msunh$.  In lower  mass clusters, the
gas may be affected significantly by  star formation and AGN activities in the
central galaxies, which gives rise to the relative strong emission in X-ray as
well as a relatively blue color.  On the other hand, for clusters with masses
above  $\sim 10^{14}\msunh$,  star formation  in member  galaxies may  be more
quenched  if  the amount  of  X-ray  gas  is larger.   Alternatively,  massive
clusters  with  stronger X-ray  emission  may  be  more relaxed  systems  that
assembled  earlier and  so their  satellite galaxies  formed earlier  and also
experienced longer time of star formation quenching.  

To  check the  reliablity of  our results,  we carry  out the  following test.
Since our halo  masses are based on the ranks of  $M_{\rm st}$, the separation
in  $\Lx$ of  clusters in  a given  halo  mass ($M_h$)  range is  the same  as
separation in a given $M_{\rm st}$ range.   To check if and to what extent the
findings of the red fraction  described above are affected by uncertainties in
the halo  masses, let  us consider the  red fractions  as a function  of $\Lx$
instead  of $M_h$.  To  this end,  we first  rank and  separate our  473 X-ray
clusters  with $S/N>3.0$  (and 1193  clusters with  $S/N>2.0$) into  six $\Lx$
bins,  each  of  which  has  exactly  the  same  number  of  clusters  as  the
corresponding $M_h$ bin.  The clusters in each $\Lx$ bin are then divided into
two subsamples, `optically strong' and `optically weak' according to the value
of $M_{\rm  st}$, so  that the  number of clusters  in the  `optically strong'
(`optically weak') subsample is the same as that in the `X-ray strong' (`X-ray
weak) subsample.  The corresponding results  of the red fractions are shown in
Fig.~\ref{fig:red_v} using dotted and dashed lines.  Here for clarity, the red
fraction  is shifted  downwards  by a  factor of  0.2.   As one  can see,  for
satellite galaxies,  the red  fraction increases with  $\Lx$, but for  a given
$\Lx$  it does  not show  any  significant difference  between the  `optically
strong' and `optically weak' subsamples.   This is consistent with the results
that  satellite galaxies  in  `X-ray strong'  clusters  of a  given $M_h$  (or
$M_{\rm st}$) on  average have higher red fractions, and  with the result that
in clusters with $M_h> 10^{14}\msunh$ the red fraction is quite independent of
$M_h$  for  both  `X-ray  strong'  and `X-ray  weak'  clusters.   For  central
galaxies, on  the other hand, here  we can see  the red fraction in  low $\Lx$
clusters does  show significant difference between the  `optically strong' and
`optically weak'  subsamples, in the  sense that, for  a given $\Lx$,  the red
fraction is higher for `optically strong' clusters. The lower red fraction (or
high blue fraction) of central galaxies in `optically weak' clusters indicates
that their relatively strong X-ray emission relative to their $M_{\rm st}$ may
be due to  high level of star formation, again in  agreement with result that,
for  a given  $M_h$ (or  $M_{\rm  st}$) in  the low-mass  end, `X-ray  strong'
clusters on average have lower red fraction (higher blue fraction) than `X-ray
weak'.   With all these  tests, we  believe our  results about  the connection
between star formation and X-ray strength is reliable.

Finally, we look at the conditional luminosity functions (CLF; see Yang, Mo \&
van den Bosch 2003) for these two subsamples of clusters.  For this purpose we
first divide  the X-ray clusters  into four mass  bins.  For each mass  bin we
determine the CLF using the same method as outlined in Yang \etal (2008).  The
results are shown  in Fig.~\ref{fig:CLF} as symbols with  error bars for X-ray
strong (filled for centrals, open  for satellites) and as histograms for X-ray
weak subsamples, respectively. Here the error bars have been obtained from 200
bootstrap  re-samplings of  all the  clusters in  question.   Different panels
correspond to different halo masses, as indicated by the value of $M_h$. Since
the halo  masses of clusters are  estimated using $M_{\rm st}$  for all member
galaxies with  $\rmag \le -19.5$, the  CLFs between the two  samples at $\rmag
\le -19.5$ are similar in both the central and satellite components.  However,
at the  fainter end,  a significant difference  between the two  subsamples is
apparent in  halos with  masses $\ga 10^{14}\msunh$:  clusters that  are X-ray
strong on average have more satellites than clusters of the same mass that are
X-ray weak.  In smaller halos with mass $\la 10^{14}\msunh$, such a difference
disappears.   We  have also  checked  the CLFs  separately  for  red and  blue
galaxies   according  to  the   color  separation   criterion  given   by  Eq.
(\ref{quadfit}). We  find that  the difference in  the satellite  CLFs between
X-ray strong  and weak  massive clusters is  mainly due  to the excess  of red
galaxies in X-strong clusters, while the blue satellite galaxies have CLF that
is  quite independent  of X-ray  emission. This  suggests that  the  excess of
satellite  galaxies  in  X-ray  strong  clusters is  mainly  due  to  survived
satellites  that were  accreted  earlier,  rather than  a  larger fraction  of
galaxies transformed from blue to red  galaxies. We have also checked the CLFs
in different $\Lx$ bins  separately for`optically strong' and `optically weak'
subsample,  and  found that  `optically  strong'  clusters have  significantly
higher   CLFs  than   `optically  weak'   clusters.   This   is   expected  by
definition. We also  found that the CLFs for the three  highest $L_X$ bins are
very similar.

\section{Conclusions}
\label{sec_summ}

Galaxy  clusters are  the largest  known gravitationally  bound  objects. With
their  combined  X-ray and  optical  properties  (e.g.  luminosities,  masses,
colors,  etc.) understood,  their  power as  cosmological  constraints can  be
significantly increased.   In addition, one  can also take advantage  of these
combined  properties to gain  insight into  the evolution  of galaxies  in the
densest regions in the Universe.

Using  an OTX  code we  developed and  the ROSAT  broadband  ($0.1$-$2.4$ keV)
archive, we have  measured the X-ray luminosities around  $\sim 65000$ optical
clusters  with masses  $\ga 10^{13}\msunh$  identified from  SDSS  DR7.  The
optical information, such as the MMG positions, halo masses, are used in X-ray
detections, which enables us to  measure X-ray luminosities more reliably than
without  such information,  as is  shown  by comparisons  with X-ray  clusters
available from the  literature.  Among these clusters, 817  have $S/N>3$ X-ray
detections, 12629 have  $3\ge S/N>1$ and 21076 have  $1\ge S/N>0$. Compared to
the 201  entries available  in the literature  from the  RASS in the  SDSS DR7
coverage,  our  817 clusters  with  $S/N>3$  already  increase the  number  of
detections by a factor of about 4.

Based on  the 473 clusters with  $S/N>3$ (sample {\bf II})  and 8780 clusters
with $S/N>0$  (sample {\bf I})  at redshift $z\le  0.12$, we have  carried out
some general analyses about the correlation between the X-ray luminosities and
various optical properties of the clusters. Our main results are summarized as
follows.
\begin{enumerate}
\item Among our X-ray clusters,  about $65\%$ of the central galaxies, defined
  to be the galaxies nearest to the X-ray flux peak, are the MMGs in clusters.
  In the remaining $\sim 35\%$, the  MMGs roughly follows a NFW profile with a
  concentration $c=6$ around the X-ray peaks.
\item The  cluster X-ray luminosity  shows correlation with the  total stellar
  mass  (or luminosity)  of the  clusters, and  with the  stellar mass  of the
  central galaxy,  but the scatter is  quite large.  The  scaling relations we
  found  are  roughly  at:  $\log  \Lx =  1.66^{+0.06}_{-0.17}~  (\log  M_{\rm
    st}-12.56^{+0.08}_{-0.03})$; $\log \Lx = 1.59^{+0.16}_{-0.08}~(\log L_{\rm
    G}  -  12.17^{+0.04}_{-0.07})$;  $\log  \Lx  =  2.46^{+0.18}_{-0.15}~(\log
  M_{\ast,  c}  - 11.75^{+0.03}_{-0.03})$;  $\log  \Lx =  2.58^{+0.12}_{-0.11}
  ~(\log  L_c -  11.24^{+0.03}_{-0.04})$, with  scatter $\sim  0.40$  in $\log
  \Lx$.
\item The scaling relation between X-ray luminosity and halo mass we obtained,
  $\log \Lx = 1.65^{+0.04}_{-0.02}~(\log M_h - 14.70^{+0.01}_{-0.01})$, is in
  quite good agreement with those obtained by Mantz \etal (2010) and Stanek et
  al. (2006).
\item Studying  the galaxy  populations in X-ray  clusters of  similar optical
  properties but different X-ray luminosities  we found that, in massive halos
  with  masses $\ga  10^{14}\msunh$,  X-ray strong  clusters  have a  larger
  fraction of  red satellite galaxies, while  the trend is  absent in relative
  lower-mass halos.
\item In relative  low mass halos with $M_h\la  10^{14}\msunh$, X-ray strong
  clusters have a smaller fraction of red central galaxies.
\item  In massive  clusters with  masses $\ga  10^{14}\msunh$,  strong X-ray
  emitters  have  many  more  low-mass  satellite  galaxies  than  weak  X-ray
  emitters. Such a difference is absence in lower mass clusters.
\item The excess of these low-mass satellite galaxies in X-ray strong clusters
  is  mainly due  to red  galaxies, suggesting  that the  excess  of satellite
  galaxies in X-ray strong clusters  is mainly due to survived satellites that
  were  accreted earlier, rather  than due  to a  larger fraction  of galaxies
  transformed from blue to red galaxies.
\end{enumerate}

\section*{Acknowledgements}

We thank the anonymous referee  for helpful comments that greatly improved the
presentation of this paper.  This work  is supported by grants from NSFC (Nos.
10925314,  11128306,  11121062,  11233005)  and  the  CAS/SAFEA  International
Partnership  Program for  Creative Research  Teams (KJCX2-YW-T23).   HJM would
like to acknowledge the support of NSF AST-0908334.

\label{lastpage}

\end{document}